\documentclass[journal,comsoc,onecolumn]{IEEEtran}
\usepackage[T1]{fontenc}

\ifCLASSINFOpdf
\else
\fi
\usepackage{amsmath}
\usepackage[cmintegrals]{newtxmath}
\usepackage[utf8]{inputenc}
\usepackage{algpseudocode}
\usepackage{subfig}
\usepackage{graphicx} 
\usepackage{array}
\usepackage{makecell}
\usepackage{eqparbox}
\usepackage{blindtext}
\usepackage{hyperref}
\usepackage{multirow}
\usepackage{listings}
\usepackage{enumitem}
\usepackage{multirow}
\usepackage{amsmath}
\usepackage{graphicx}
\usepackage{amssymb}
\usepackage[english]{babel}
\usepackage{multirow}
\usepackage{booktabs}
\usepackage{mathtools}
\usepackage{nameref}
\usepackage{relsize}
\usepackage[usenames, dvipsnames]{color}
\usepackage{booktabs}
\newcommand{\tabitem}{~~\llap{\textbullet}~~}
\usepackage{enumitem}
\usepackage{lineno}
\usepackage{stackengine}
\begin{document}
%
\title{ECG Language Processing (ELP): a New Technique to Analyze ECG Signals}
%
%
%

\author{Sajad~Mousavi,
        Fatemeh~Afghah,
        Fatemeh~Khadem,
        and~U. Rajendra~Acharya
\thanks{S. Mousavi and F. Afghah are with the School of Informatics, Computing, and Cyber Systems, Northern Arizona University, Flagstaff, AZ, 86011 USA (e-mail: SajadMousavi@nau.edu, Fatemeh.Afghah@nau.edu).}
\thanks{F. Khadem is an independent researcher (e-mail: Fatemeh.Khadem019@gmail.com).}
\thanks{U. R.~Acharya is with the School of Engineering, Ngee Ann Polytechnic, Singapore, the School of Science and Technology, Singapore University of Social Sciences, 463 Clementi Road, 599494, Singapore, and the Department Bioinformatics and Medical Engineering, Asia University, Taiwan (e-mail: aru@np.edu.sg).}
}

\maketitle

\begin{abstract}
A language is constructed of a finite/infinite set of sentences composing of words. Similar to natural languages, Electrocardiogram (ECG) signal, the most common noninvasive tool to study the functionality of heart and diagnose several abnormal arrhythmia, is made up of sequences of three or four distinct waves including the P-wave, QRS complex, T-wave and U-wave. An ECG signal may contain several different varieties of each wave (e.g., the QRS complex can have various appearances). For this reason, the ECG signal is a sequence of heartbeats similar to  sentences in natural languages) and each heartbeat is composed of a set of waves (similar to words in a sentence) of different morphologies. Analogous to natural language processing (NLP) which is used to help computers understand and interpret the human's natural language, it is possible to develop methods inspired by NLP to aid computers to gain a deeper understanding of Electrocardiogram signals. In this work, our goal is to propose a novel ECG analysis technique, \textit{ECG language processing (ELP)}, focusing on empowering computers to understand ECG signals in a way physicians do. We evaluated the proposed method on two tasks including the classification of heartbeats and the detection of atrial fibrillation in the ECG signals. Experimental results on three databases (i.e., PhysionNet's MIT-BIH, MIT-BIH AFIB and PhysioNet Challenge 2017 AFIB Dataset databases) reveal that the proposed method is a general idea that can be applied to a variety of biomedical applications and is able to achieve remarkable performance.
\end{abstract}


%
\IEEEpeerreviewmaketitle


\section{Introduction}
ECG is the most common signal used by physicians and cardiologists to monitor the functionality of the heart. Manual analysis of ECG signals by a human is a very challenging and time-consuming task due to dealing with long ECG recordings and the existence of complex patterns associated with different heart arrhythmia in the ECG signal. Therefore, to deal with the issues related to the manual analysis of ECG signals, several studies focus on developing  automatic ECG analysis techniques to perform this task with high accuracy and in a real-time manner. Machine learning algorithms are commonly used to detect the arrhythmia in the ECG signals \cite{acharya2017deep,kachuee2018ecg,mousavi2020single,zaeri2018feature}. Typically, these methods consider four main steps in their workflows: (1) Pre-possessing signal that includes re-sampling the signals, noise removal (using band-pass filters, etc.), signal normalization/standardization, etc., (2) Heartbeat segmentation that involves detection of the R-peak (i.e., the QRS complex) using some algorithms such as Pan and Tompkins's algorithm \cite{pan1985real}, open-source gqrs package provided by Physioent community \cite{PhysioNet}, etc., (3) Feature extraction that includes transforming raw signal to features best suited to the specific task (i.e., classification, prediction, regression, etc.). and (4) Learning that considers classical machine learning techniques such as  multilayer perceptron (MLP) and decision trees for analysing ECG signals \cite{zaeri2018feature}. 

Even though conventional machine learning algorithms with the handcrafted features have achieved acceptable performance for ECG analysis, deep learning models with the power of automated feature extraction and representation learning have proven to get human-level performance in analyzing biomedical signals \cite{rajpurkar2017cardiologist,yildirim2018arrhythmia,murat2020application}. However, deep learning techniques need a large amount of data and are composed of huge parameters to be learned. In addition, most of the suggested methods and workflows for analyzing ECG signals are tailored to the specific task and are not generalizable to other biomedical problems.

\begin{figure*}[htb]
\centering
  \includegraphics[height=0.1\textheight,width=1.\linewidth]{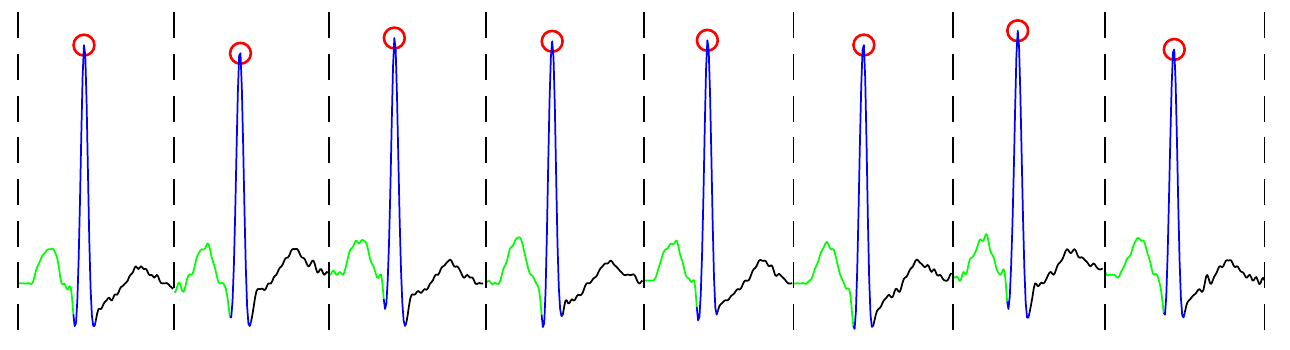}
  \caption{Illustration of an ECG signal; Red circles indicate R peaks; green, blue and black curves illustrate P, QRS and T waves respectively.} 
  \label{fig:ecg_wave}
\end{figure*}
\begin{figure}[htb]
\centering
  \includegraphics[height=0.3\textheight,width=1.\linewidth,keepaspectratio]{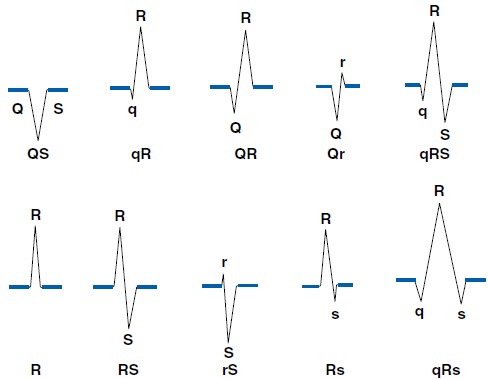}
  \caption{the QRS complex morphology; adopted from \cite{qrscomplexvary}.} 
  \label{fig:qrs_complex}
\end{figure}

In this study, we open a new research avenue for ECG signal analysis by introducing a novel framework called \textit{ECG language processing (ELP)} that processes the ECG signal in a way a text document is treated in natural language processing (NLP) framework. The proposed framework is applicable to various biomedical applications and also can improve the performance of the shallow machine learning algorithms. A language is constructed of a finite/infinite set of sentences composing of words. Similar to natural languages, an ECG signal is made up of sequences of three or four distinct waves including the P-wave, QRS complex, T-wave and U-wave \cite{hurst1998naming,ecgcycle2018} (refer to Figure \ref{fig:ecg_wave}). Each normal ECG includes different varieties of each wave. For instance, the QRS complex can have various shapes as shown in Figure \ref{fig:qrs_complex}. Hence, an ECG signal is a sequence of heartbeats (like sentences in natural languages) and each heartbeat is composed of a set of waves (like words in a sentence) of different morphologies. Analogous to NLP which is utilized to help computers/machines to understand and interpret the human's natural language, our proposed NLP-inspired ECG language processing can aid the computers to gain a deeper understanding of Electrocardiogram signals.

The rest of this work is structured as follows. Section \ref{sec:methodology_1} explains the proposed ELP method. Section \ref{sec:elpexamples} introduces potential applications of the ELP method. Section \ref{sec:experi} presents the experimental setup, the used datasets to assess the suggested method, and gives a performance comparison of the proposed approach against the existing algorithms in the literature, following by a discussion. Finally, Section \ref{sec:conclu} concludes the study.

\section{Methodology}
\label{sec:methodology_1}
In this section, we describe main components of ECG Language Processing. Figure \ref{fig:ELP-pipeline} shows the ELP pipeline. The ELP includes two main steps as follows:
\begin{figure*}[ht]
\centering
  \includegraphics[height=4cm,width=1\linewidth]{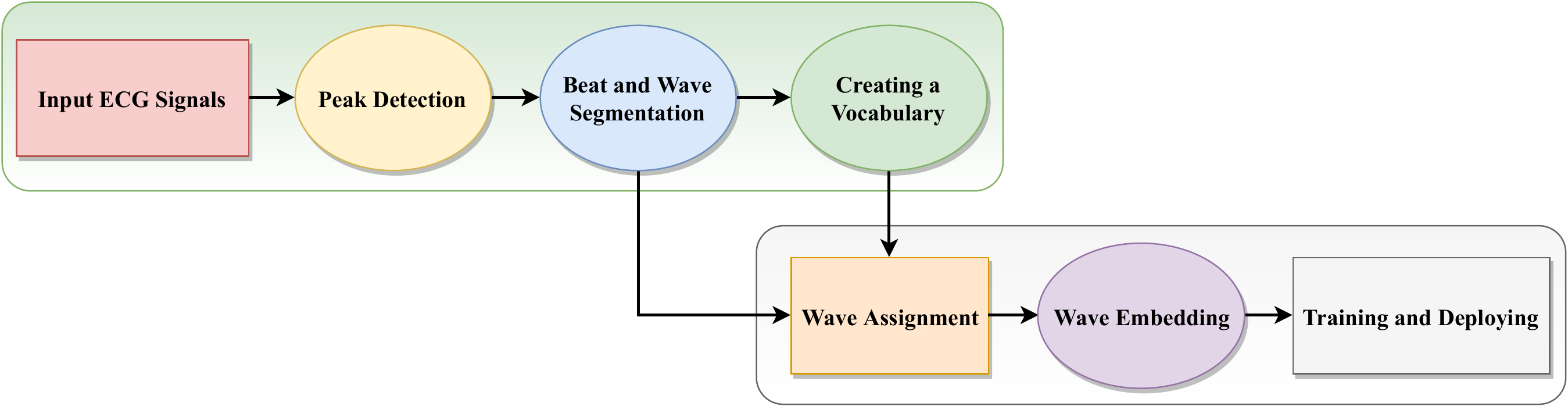}
  \caption{ECG language processing Pipeline.} 
  \label{fig:ELP-pipeline}
\end{figure*}


\subsection*{Step 1: Creating a Wave Vocabulary}
\label{step_1}
\begin{figure*}[htb]
\centering
  \includegraphics[width=1.\linewidth]{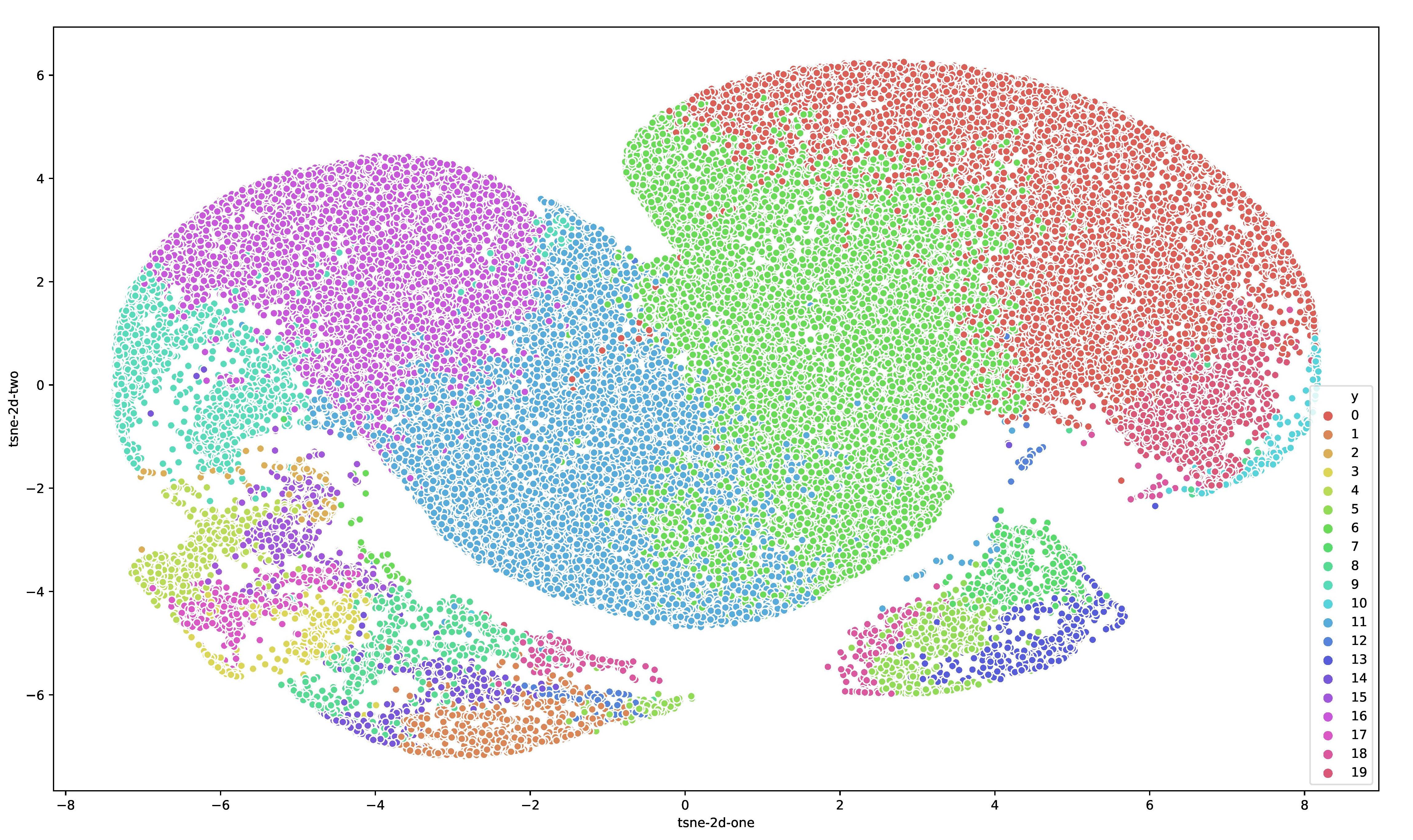}
  \caption{Visualizations of thousands of extracted waves along with their 20 clusters from the PhysioNet Computing in Cardiology Challenge 2017 dataset.} 
  \label{fig:t_SNE_1}
\end{figure*}

\begin{figure*}
\footnotesize
\includegraphics[width=0.5\linewidth,height=0.1\textheight]{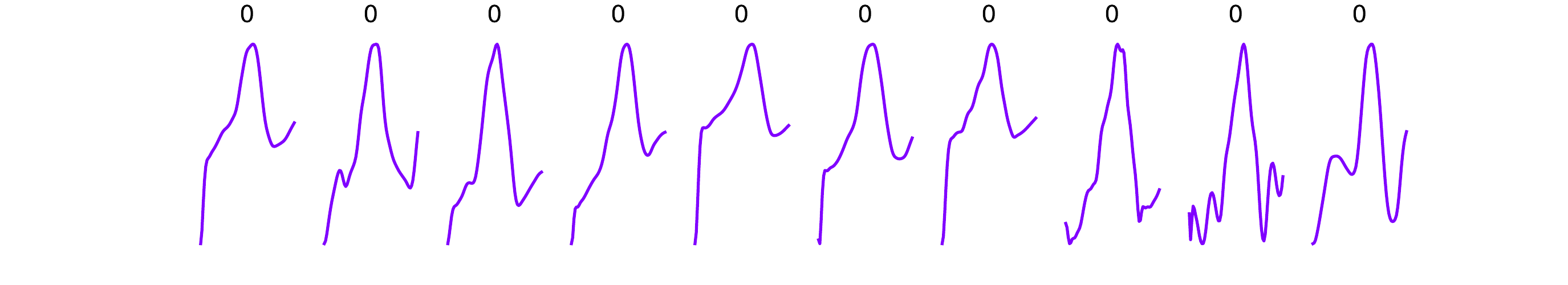}%
\includegraphics[width=0.5\linewidth,height=0.1\textheight]{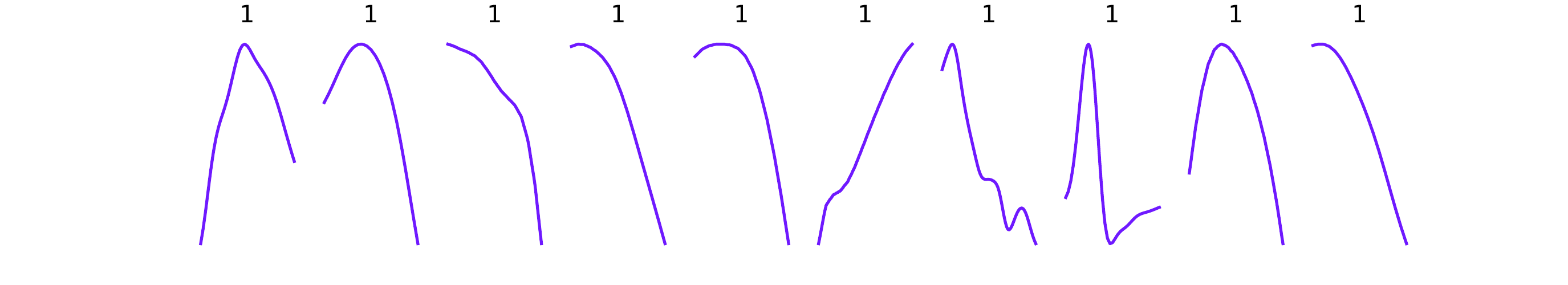}
\includegraphics[width=0.5\linewidth,height=0.1\textheight]{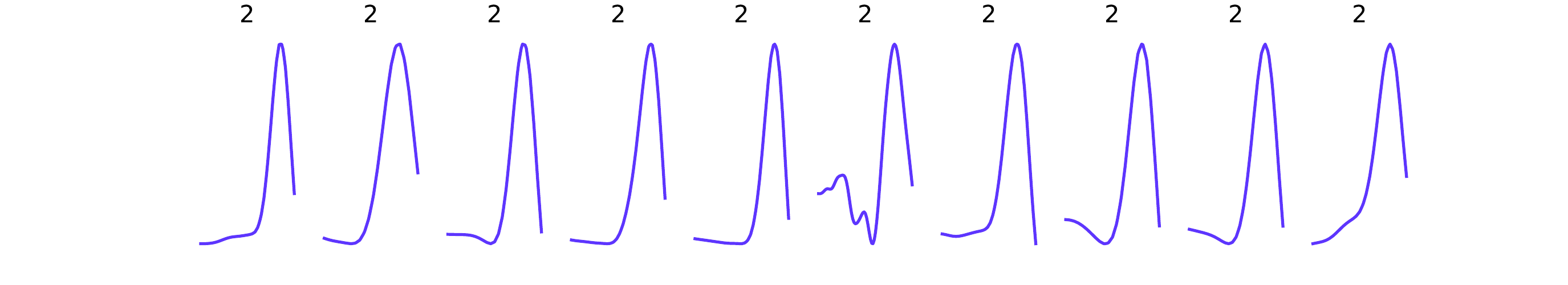}
\includegraphics[width=0.5\linewidth,height=0.1\textheight]{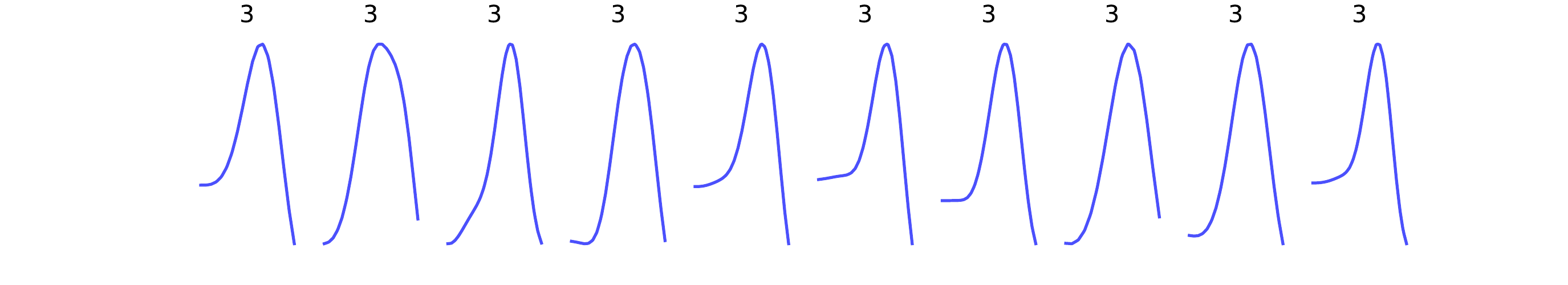}
\includegraphics[width=0.5\linewidth,height=0.1\textheight]{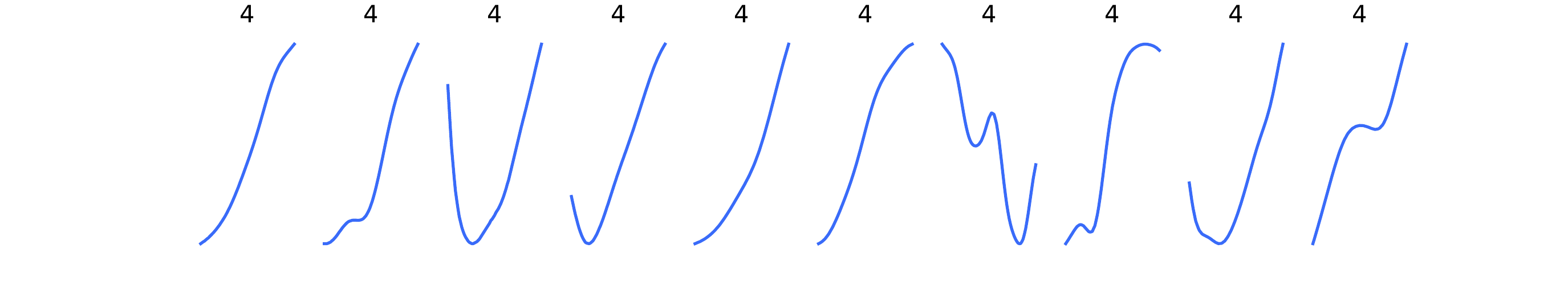}
\includegraphics[width=0.5\linewidth,height=0.1\textheight]{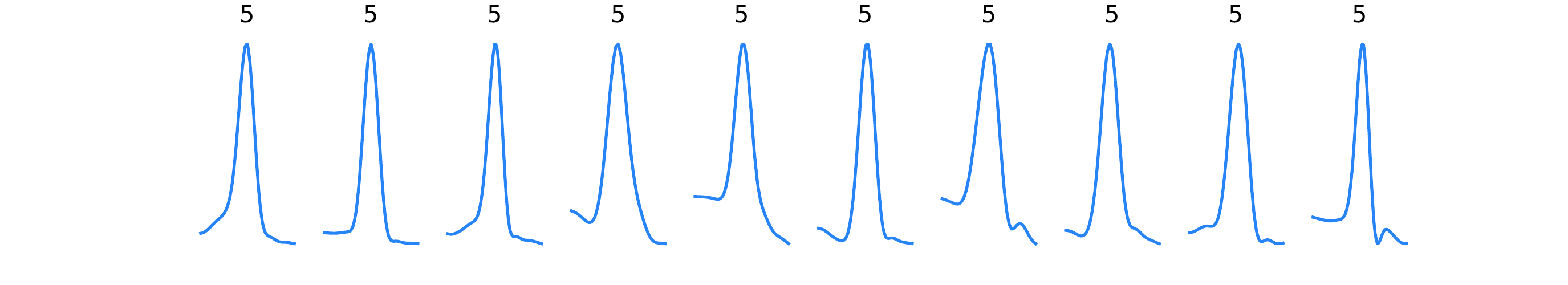}
\includegraphics[width=0.5\linewidth,height=0.1\textheight]{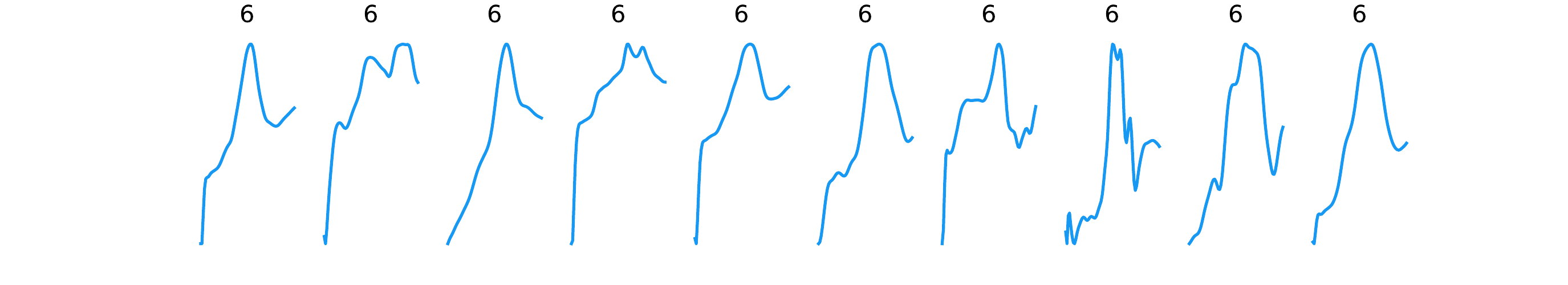}
\includegraphics[width=0.5\linewidth,height=0.1\textheight]{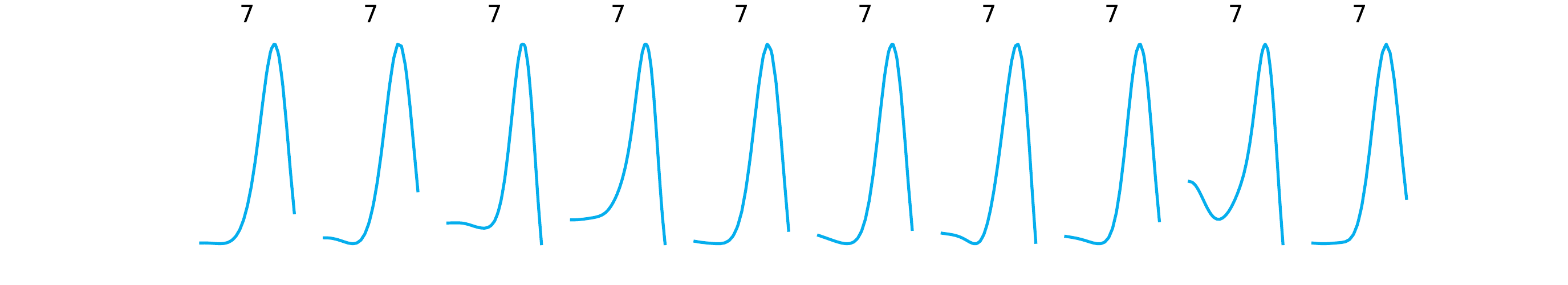}
\includegraphics[width=0.5\linewidth,height=0.1\textheight]{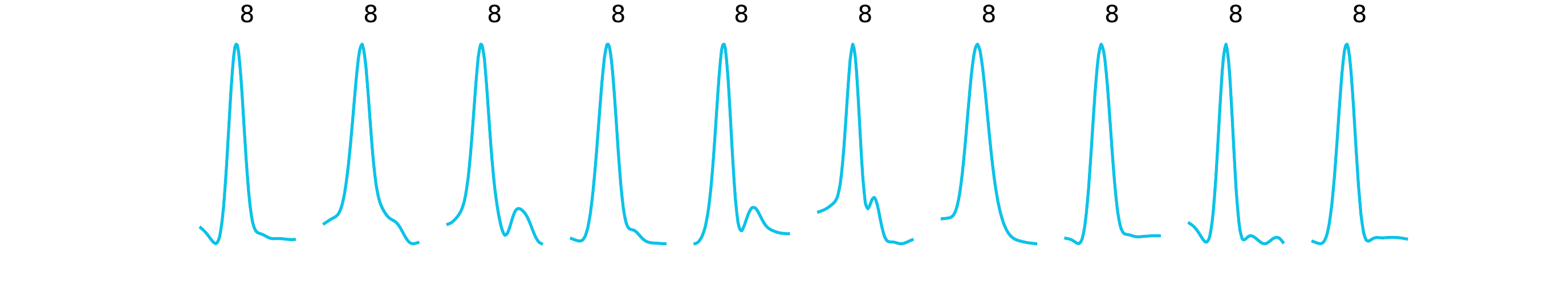}
\includegraphics[width=0.5\linewidth,height=0.1\textheight]{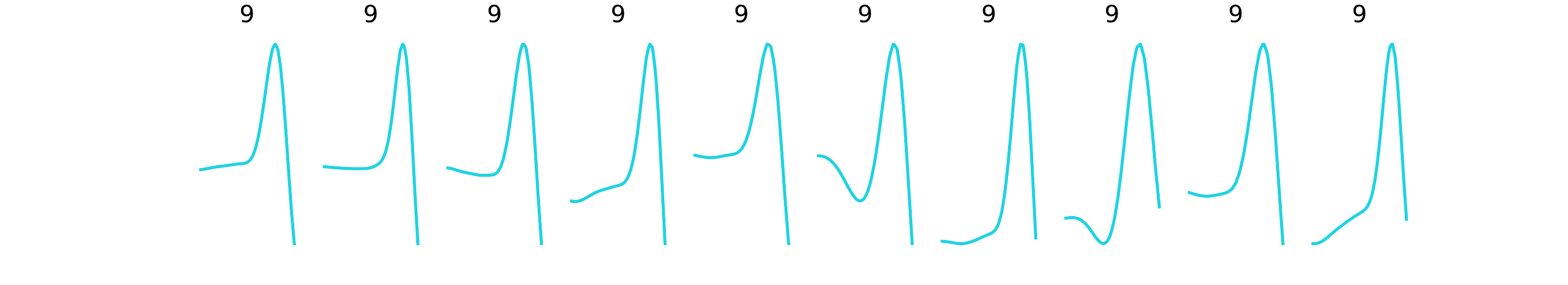}
\includegraphics[width=0.5\linewidth,height=0.1\textheight]{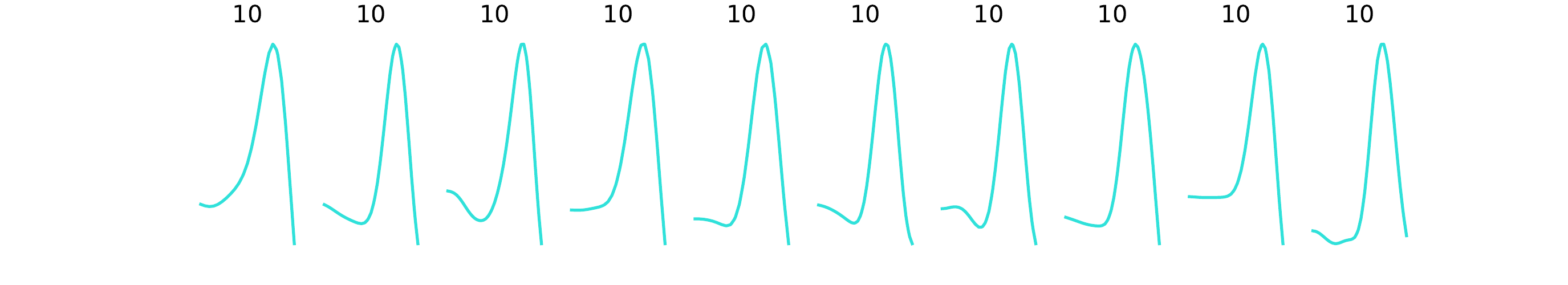}
\includegraphics[width=0.5\linewidth,height=0.1\textheight]{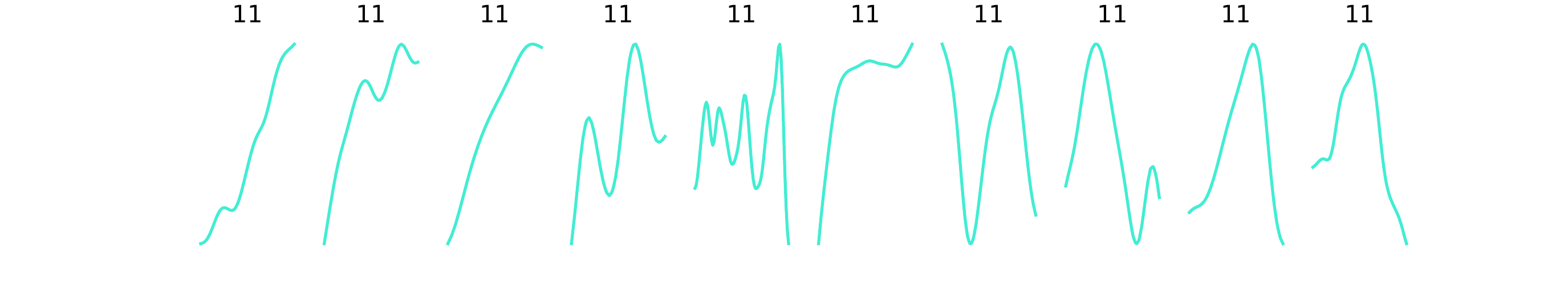}
\includegraphics[width=0.5\linewidth,height=0.1\textheight]{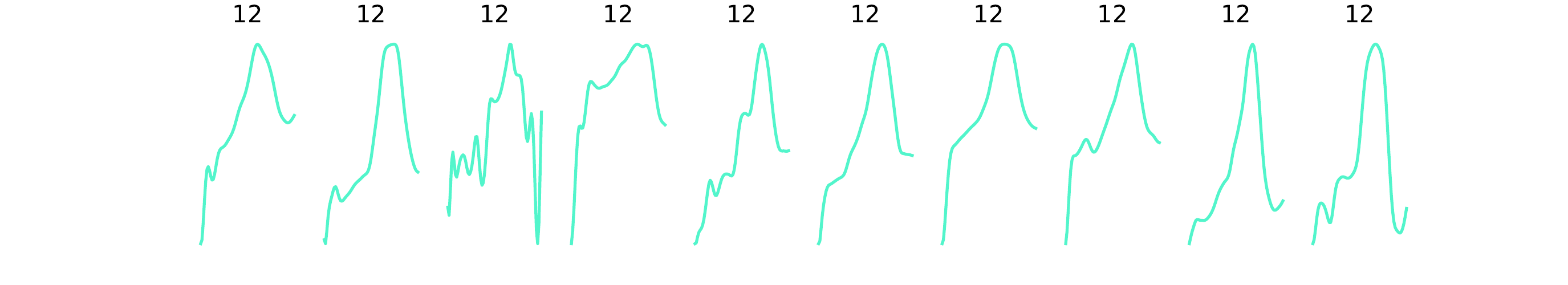}
\includegraphics[width=0.5\linewidth,height=0.1\textheight]{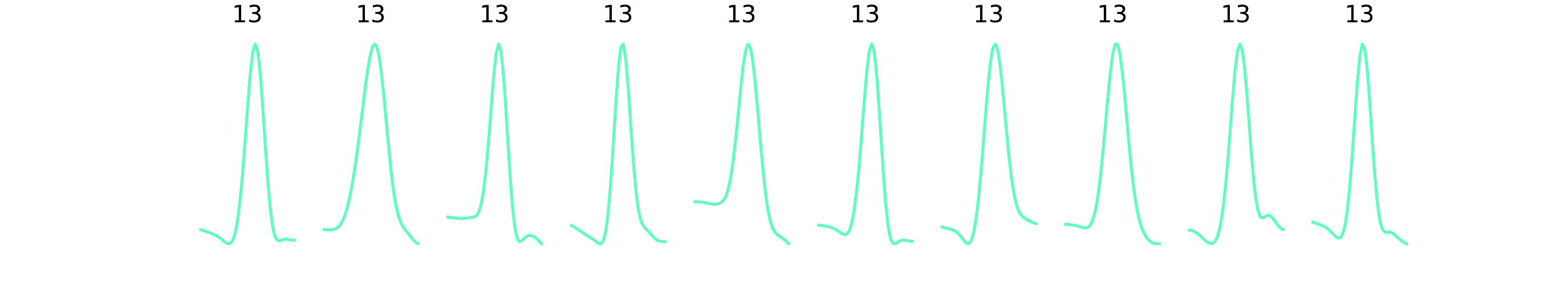}
\includegraphics[width=0.5\linewidth,height=0.1\textheight]{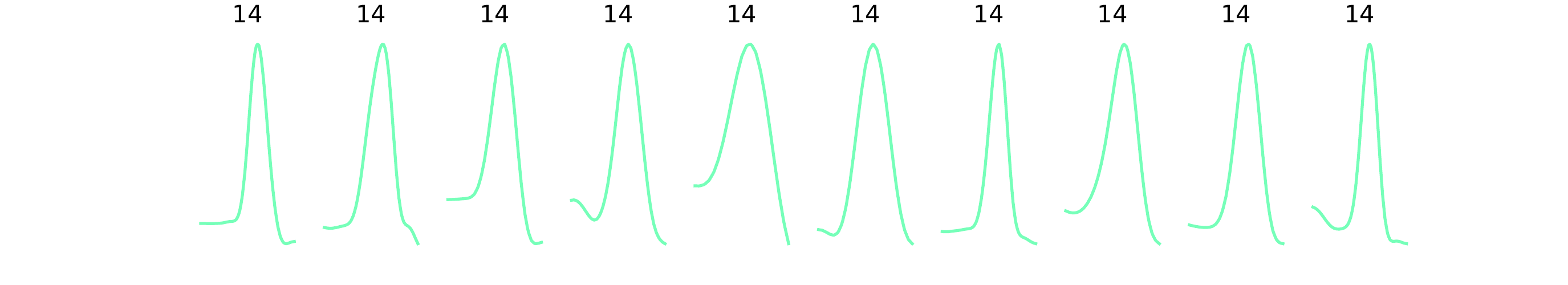}
\includegraphics[width=0.5\linewidth,height=0.1\textheight]{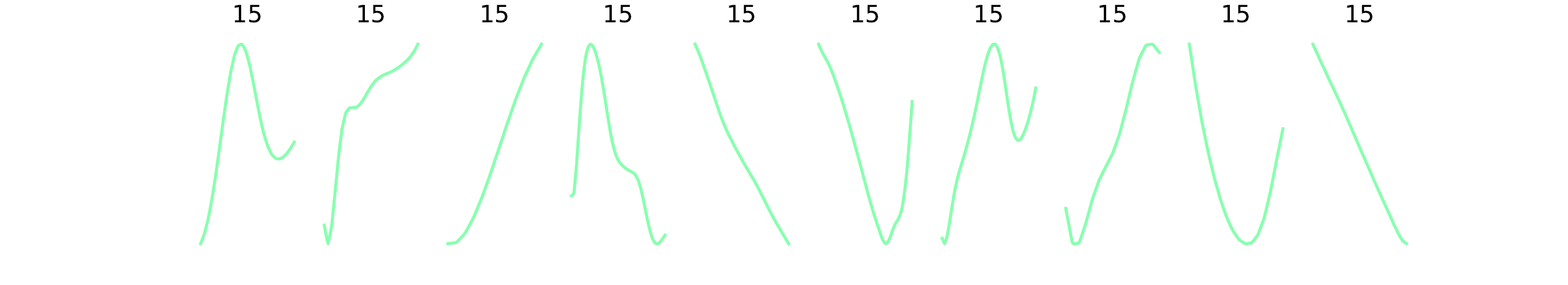}
\includegraphics[width=0.5\linewidth,height=0.1\textheight]{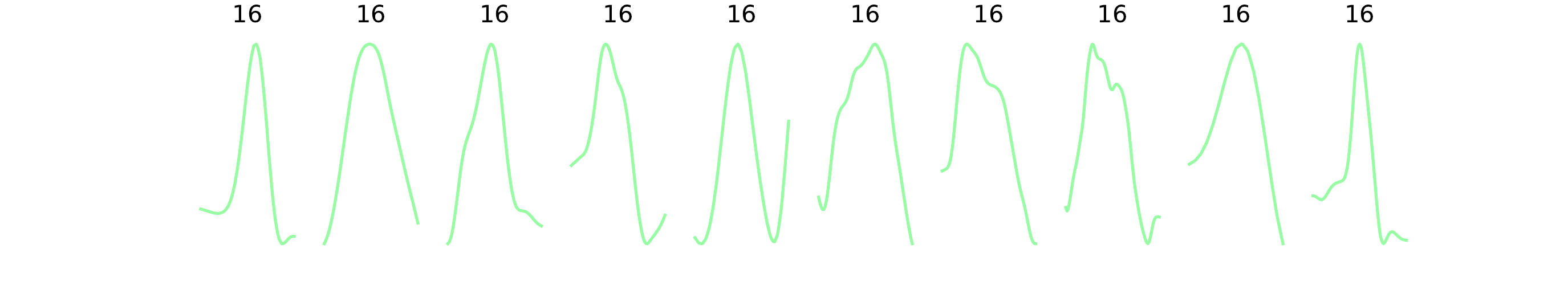}
\includegraphics[width=0.5\linewidth,height=0.1\textheight]{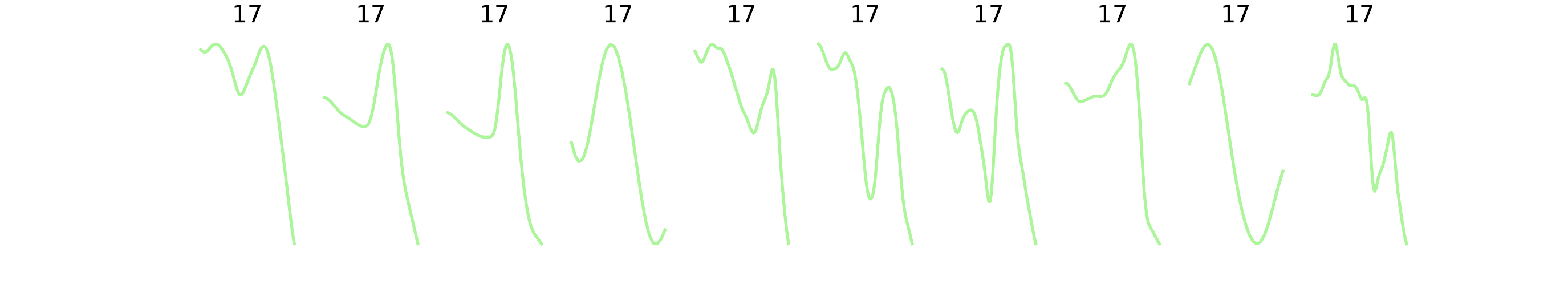}
\includegraphics[width=0.5\linewidth,height=0.1\textheight]{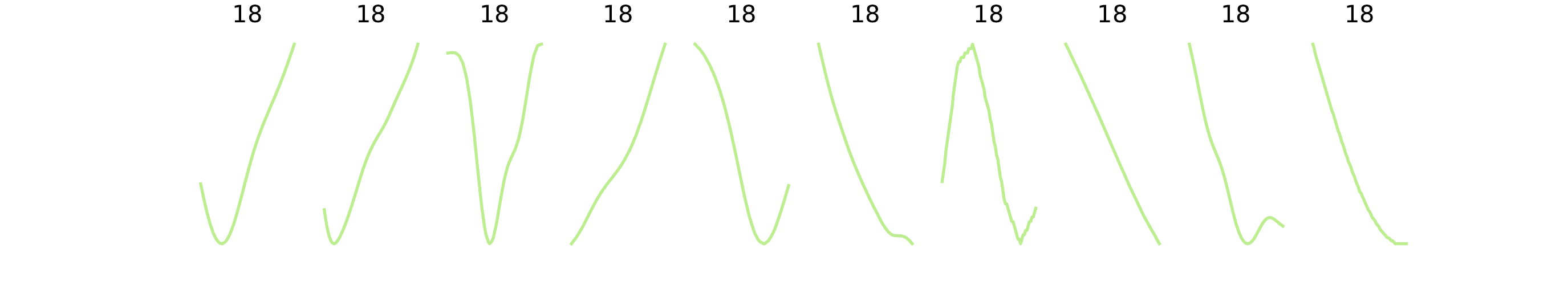}
\includegraphics[width=0.5\linewidth,height=0.1\textheight]{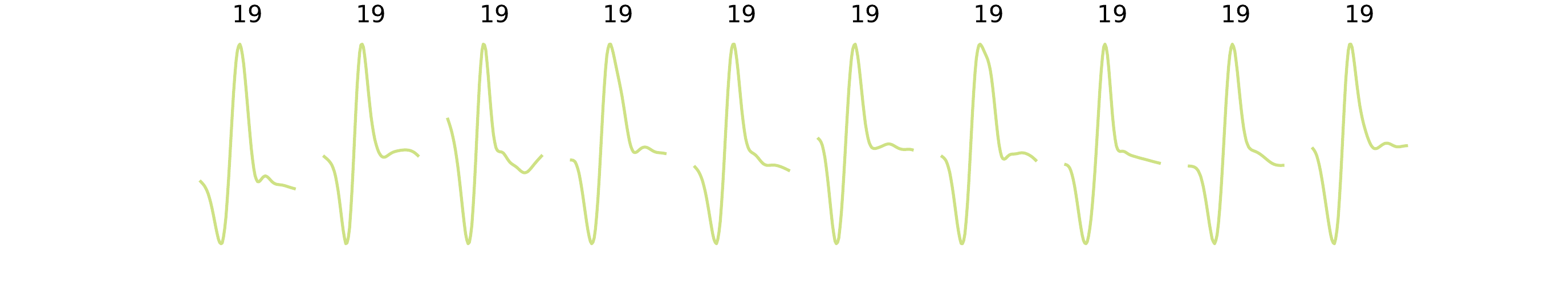}
\caption{Visualization of extracted waves using a clustering algorithm and their corresponding clusters; the numbers above waves indicate their cluster number.}
\label{fig:cluster_samps}
\end{figure*}

\begin{itemize}

\item \textbf{Peak Detection:} it includes detecting the R-peaks of given ECG signal or detecting the QRS complexes. The commonly used algorithms for such purpose are the Pan–Tompkins algorithm \cite{pan1985real, sathyapriya2014analysis} and one proposed by \cite{manikandan2012novel}. The red circles in Figure \ref{fig:ecg_wave} depicts the R-peaks of a sample ECG signal. 

\item \textbf{Beat and Wave Segmentation:} it involves dividing continuous ECG signal into a sequence of heartbeats, and split the heartbeats into distinct units called waves. After detecting R-peaks, the presence of other building waves (i.e., P, QRS and T waves) in the ECG signal can be extracted using adaptive searching windows. To do heartbeat segmentation, one can identify a segment as a fixed number of samples before  the R-peak location to the fixed number of samples after the R-peak location or from the onset of the P-wave to the offset of consecutive T-wave. Figure \ref{fig:ecg_wave} depicts a segmented ECG signal annotated with the R-peaks, P, QRS and T waves. 

\item \textbf{Creating a Vocabulary:} it includes building a vocabulary of the waves based on the extracted waves from the ECG signals. We can cluster all the waves, then consider the mean of each cluster as an entry of the vocabulary. This can be done by feeding all waves into off-the-shelf  clustering algorithms such as K-means, spectral clustering or agglomerative clustering algorithms \cite{kanungo2002efficient,xu2005survey, ng2002spectral}. After doing wave clustering, the mean of each cluster can represent a distinct wave of the vocabulary. Figure \ref{fig:t_SNE_1} visualizes the extracted waves of an ECG signal dataset and extracted clusters (20 clusters) using t-Distributed Stochastic Neighbor Embedding (t-SNE) technique \cite{maaten2008visualizing}. Figure \ref{fig:cluster_samps} shows a wave clustering results on the dataset of PhysioNet Computing in Cardiology Challenge 2017 \cite{PhysioNetafdb17}. Each row of the figure presents 10 sample waves of a specific extracted cluster. 
\end{itemize}

\subsection*{Step 2: Training and Deploying}
\begin{itemize}
    \item \textbf{Wave Assignment} the beat and wave segmentation process produce a sequence of waves for each ECG signal. Then, the cluster of each wave of the sequence is identified using the output of the previous step (i.e., the step \ref{step_1} of pipeline). In other words, it assigns a unique integer value (the cluster number) to each wave of the sequence. In this way, each ECG signal is integer-encoded, so that each integer represents a specific wave (or cluster) in the vocabulary.
    \item \textbf{Wave Embedding or Wave Vectorization} it takes the integer-encoded vocabulary and builds the embedding vector (i.e., a vector of a specified length) for each wave of the vocabulary. The main reason behind word embedding is that it allows us to apply advanced machine leaning like artificial neural networks on the integer-encoded ECG signals for a specific task. Inspired by natural language processing, we can use several approaches such as Count Vectorizer in which a sequence of waves is converted into a fixed-length vector with the size of the vocabulary. The value in each position in the vector would be a count of each wave in the encoded signal, or Word2Vec approach that uses neural network techniques to represent waves in a vector space. The latter approach is more efficient so that it recognizes context, relation and similarity between waves \cite{mikolov2013distributed}.
    
    \item \textbf{Training and Deploying} it involves using machine learning and deep learning techniques to train models on output of the wave embedding step for any learning tasks including classification, prediction, etc. To give a better understanding of ELP applications, we outline some main ECG language processing examples in the following section.

\end{itemize}

\section{ECG Language Processing Examples}
\label{sec:elpexamples}
ECG Language Processing (ELP) can be used in a variety of biomedical applications where the collected data are ECG signals. Below are the most common applications of ELP:
\begin{itemize}
  \item \textbf{Heartbeat classification/detection} it involves assigning a specific label to heartbeats of a given ECG signal.  
  \item \textbf{Arrhythmia prediction} it includes predicting onset of life-threatening arrhythmia such as Atrial Fibrillation (AFib) in patients based on their current and past states. 
  \item \textbf{Automatic heartbeat annotation} it involves automatic annotation of the heartbeats in a sequence of heartbeats (i.e., an ECG signal). This problem is also called automatic sequence labeling \cite{ma2016end,mousavi2019inter}.
  \item \textbf{Summarize a long ECG signal} ECG signals are typically 8 hours (or more) long (e.g., in sleep stage detection task). Thus, interpreting such a long ECG signal by cardiologists and physicians is a very time-consuming and prone to error task. One way to tackle this problem is summarizing the ECG signal and only extracting the most related regions of the ECG signal  which contribute  to a specific event. This can be done by using attention mechanisms \cite{xu2015show,mousavi2019ecgnet}.
  \item \textbf{Translate ECG to other physiological signals} it involves estimating other physiological signals such as Arterial Blood Pressure (ABP) and photoplethysmography (PPG) from ECG signals. The main application of such translations is imputation in which the missing values (may be caused by the device inadvertent detachment) of the signals can be estimated.
\end{itemize}

\section{Experiments}
\label{sec:experi}
In this section, we evaluate our proposed ECG analysis approach (i.e., ECG Language Processing) using two different clinical tasks including atrial fibrillation detection and automatic heartbeat classification. We show performing the ELP pipeline to process ECG signals results in better performance compared to the existing methods.
\subsection{Data Description}
Two datasets including the MIT-BIH AFIB database \cite{PhysioNetmitafdb} and the PhysioNet Computing in Cardiology Challenge 2017 dataset  \cite{PhysioNetafdb17} were utilized to build models to perform the detection of atrial fibrillation, and the PhysioNet MIT-BIH Arrhythmia database \cite{PhysioNetmitdb} was used to build an automatic heartbeat annotation model.

\textbf{MIT-BIH AFIB Dataset:} This dataset includes 23 long-term ECG recordings of subjects with mostly AFIB arrhythmia. Each subject of the MIT-BIH AFIB contains two 10-hours long ECG recordings (ECG1 and ECG2). The ECG recordings are sampled at 250 Hz with 12-bit resolution over a range of $\pm10$ millivolts. In this paper, we split each ECG signal into 5-s data segments and annotated each one based on a threshold parameter, $p$. We considered the labeling method used by \cite{xia2018detecting,asgari2015automatic}. Indeed, a 5-s data segment is considered as AFIB if the percentage of labeled AFIB heartbeats of the segment is greater than or equal to $p$, otherwise it is labeled as a non-AFIB arrhythmia. Similar to the literature, the parameter $p$ was set to $50\%$. 
We extracted a total of $167,422$ 5-s data segments from the ECG1 recordings of the dataset. The number of AFIB and non-AFIB samples were $66, 939$ and $100, 483$, respectively. To cope with the class imbalance problem existing in the extracted data segments, we randomly selected the same number of segments for both AFIB and non-AFIB classes in which we considered 66, 939 data segments for both classes.

\textbf{PhysioNet Challenge AFIB Dataset:} This dataset  was applied for the PhysioNet Challenge 2017 in which the purpose was to propose algorithms to classify a single-short-ECG lead recording (with duration 30-60s) to normal sinus rhythm (N), atrial fibrillation (AFIB), an alternative rhythm (O), or too noisy ($\sim$) classes. The training set contains 8,528 single lead ECG recordings and the test set includes 3,658 ECG recordings. Because the test set has not been publicly available, we utilized the training set for building and evaluating the model. The ECG recordings were recorded by AliveCor devices, sampled as 300 Hz and filtered by a band pass filter. Table \ref{tab:StatPHyAFIBDataset} shows the statistics of the numbers of each classification type in the PhysioNet Challenge AFIB database (i.e., the training set).
\begin{table}[!ht]
\centering{
    \caption{ Details of number of each classification type in the PhysioNet Challenge AFIB dataset.}
    \label{tab:StatPHyAFIBDataset} 
    \resizebox{0.8\linewidth}{!}{ 
    \begin{tabular}{l|c|c|c|c|c} 
    \toprule[\heavyrulewidth]
  
      \textbf{Dataset} & \textbf{N} & \textbf{AFIB} &  \textbf{O} &\textbf{$\sim$}  & \textbf{Total}\\ 
       \hline
       PhysioNet Challenge AFIB & 5,154 & 771 & 2,557&46&8,528\\
     
      \bottomrule
    \end{tabular}
    }
}
\end{table}


\textbf{PhysioNet MIT-BIH:} This arrhythmia's dataset contains the ECG signals for 48 different subjects. The signals were recorded at the sampling rate of 360Hz, and each record includes two ECG leads; ECG lead II and lead V1. In this study, to be consistent with the previous works in the literature, the ECG lead II is used to build the heartbeat annotator. The dataset is recommended by the American association of medical instrumentation (AAMI) \cite{ec571998testing} and is composed of the five essential arrhythmia groups. Table \ref{tab:cat_beats} presents the categories of heartbeats existed in the database and Table \ref{tab:StatPHyMITBIHDataset} shows the statistics of the numbers of each heartbeat group in the MIT-BIH database.

\begin{table}[h]  
\caption{Groups of heartbeats presented in the
MIT-BIH database based on AAMI.}
 \centering{
\label{tab:cat_beats}
	\resizebox{0.7\linewidth}{!}{ 
\begin{tabular}{c|l}
\toprule
\textbf{Category} &  \textbf{Class}\\
\midrule
 \multirow{4}{*}{\textbf{N}} &  \tabitem Normal beat (N)\\
 &\tabitem Left and right bundle branch block beats (L,R) \\
  &\tabitem Atrial escape beat (e) \\
    &\tabitem Nodal (junctional) escape beat (j) \\
    \\
     \multirow{4}{*}{\textbf{S}} & \tabitem Atrial premature beat (A)\\
 &\tabitem Aberrated atrial premature beat (a) \\
  &\tabitem Nodal (junctional) premature beat (J) \\
    &\tabitem Supraventricular premature beat (S) \\
    
        \\
     \multirow{2}{*}{\textbf{V}} & \tabitem Premature ventricular contraction (V)\\
 &\tabitem Ventricular escape beat (E) \\
\\
\multirow{2}{*}{\textbf{F}} & \tabitem Fusion of ventricular and normal beat (F)\\
\\
\multirow{3}{*}{\textbf{Q}} & \tabitem Paced beat (/)\\
 & \tabitem Fusion of paced and normal beat (f)\\
  & \tabitem Unclassifiable beat (U)\\

 \bottomrule  
\end{tabular} }
}
\end{table}

\begin{table}[!ht]
\centering{
    \caption{Details of number of each heartbeat group in the MIT-BIH database.}
    \label{tab:StatPHyMITBIHDataset} 
    \resizebox{0.8\linewidth}{!}{ 
    \begin{tabular}{l|c|c|c|c|c|c} 
    \toprule[\heavyrulewidth]
  
      \textbf{Dataset} & \textbf{N} & \textbf{S} &  \textbf{V} &\textbf{F} &\textbf{Q} & \textbf{Total}\\ 
       \hline
       MIT-BIH Arrhythmia & 90,462 & 2,777 &  7,223& 802&8,027&109,291\\

      \bottomrule
    \end{tabular}
    }
}
\end{table}
\subsection{Experimental setup}
\label{sec:experisetup}
We built three different neural networks for each clinical task and compared them to the state-of-the-art algorithms. Below is list of the models we used to build the detective models.
\begin{figure*}[htb]%
    \centering
    \subfloat[CNN]{\includegraphics[height=11cm, width=5cm]{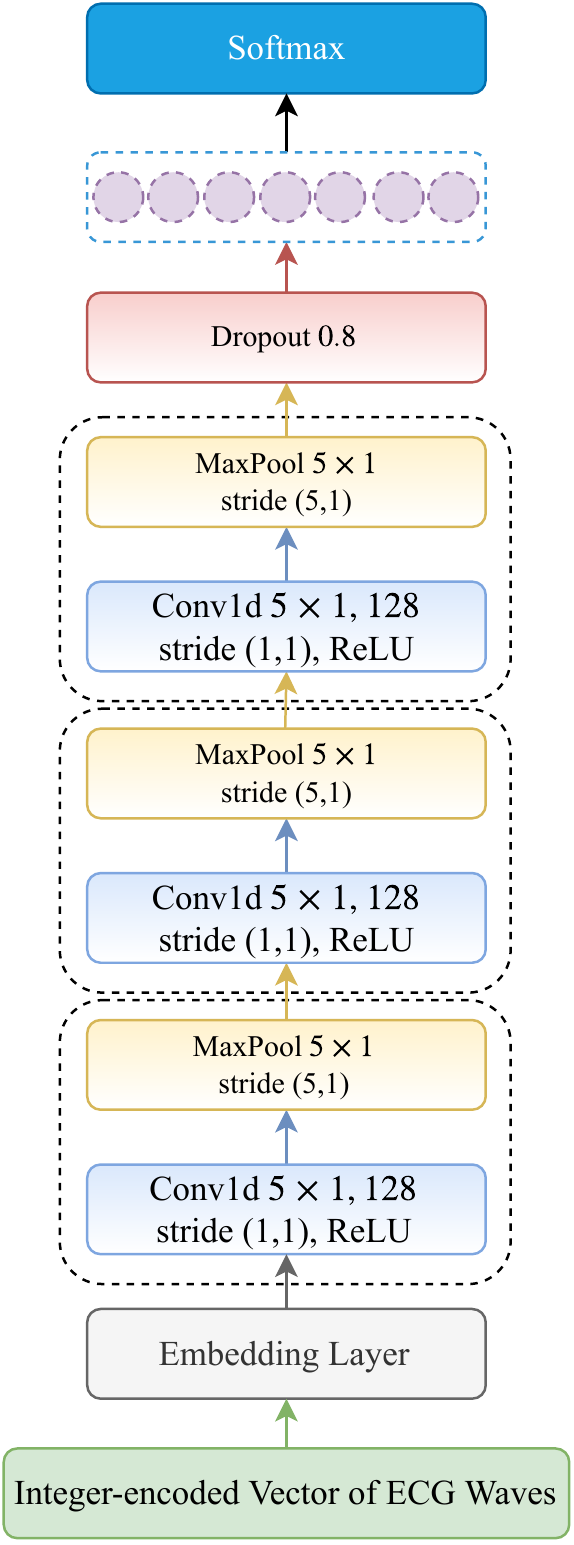} 
   \label{fig:cnn}
    }%
     \subfloat[RNN]{\includegraphics[width=5cm]{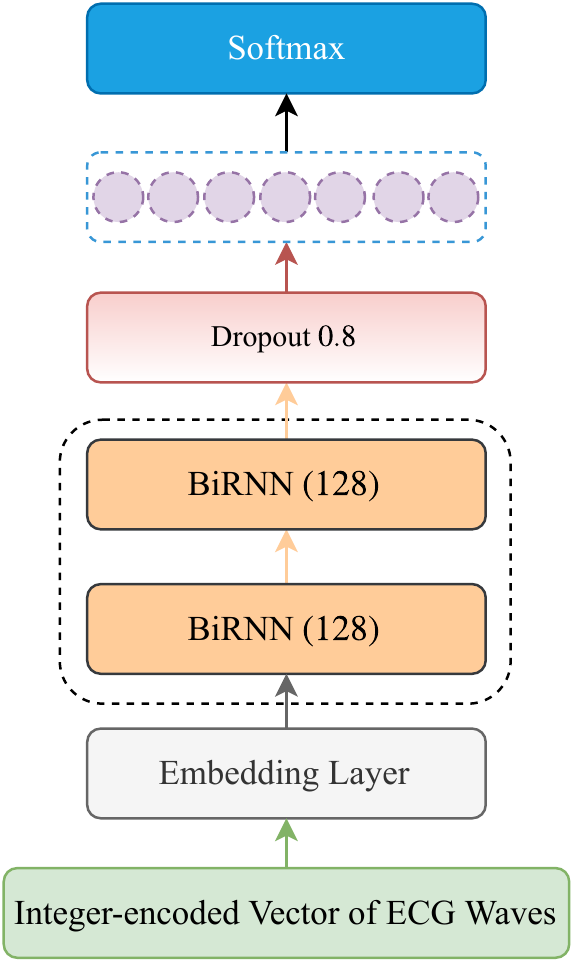}  
     \label{fig:rnn}
      }%
     \subfloat[RNN-Attention]{\includegraphics[width=5cm]{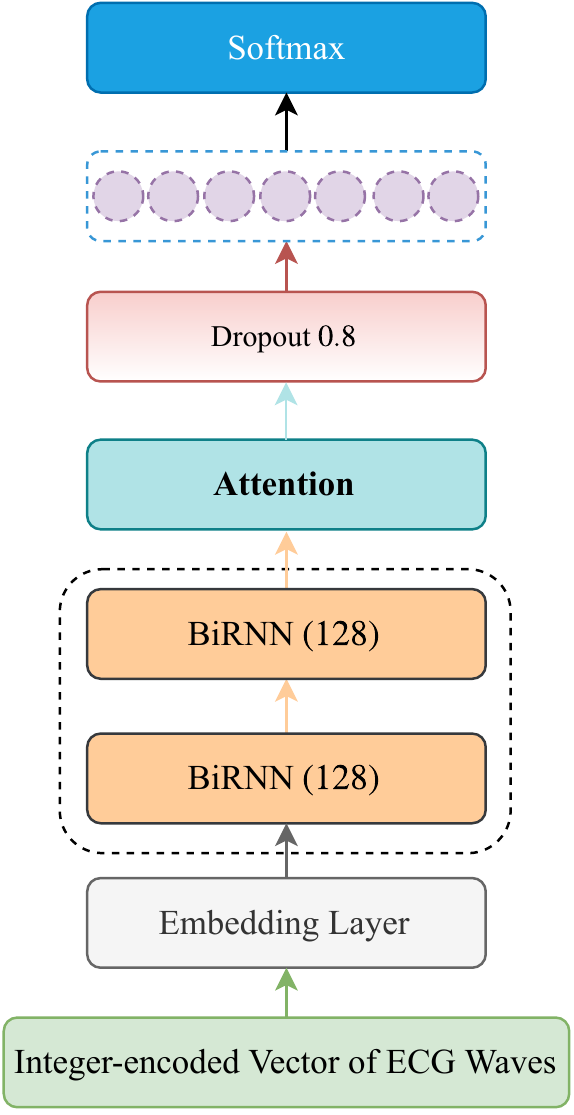}  
      \label{fig:rnn-att}
      }%
    \caption{Architectures of the used networks in the experiments.}%
    \label{fig:archs}%
\end{figure*}

\begin{itemize}
\item \textbf{Convolutional neural network (CNN)} we use three consecutive 1D convolutional layers in which each layer is composed of 128 filters with a kernel size of $5\times1$, a stride 1 and a Rectified Linear Unit (ReLU) activation function. All convolutional layers are followed by max-pooling layers with pooling regions of size $5\times1$ with stride sizes of 5. The output of the last convolutional layer is passed through a dropout layer followed by a fully-connected layer with a size of 64 followed by a softmax layer to perform the classification task  (see Figure \ref{fig:cnn}). Because the length of input signals in the MIT-BIH AFIB and PhysioNet MIT-BIH databases were too short, we use two consecutive 1D convolutional layers for both datasets with small  pooling regions of sizes $3\times1$ with a stride 3 and $2\times1$ with a stride 2, respectfully.

\item \textbf{Recurrent neural network (RNN)} we utilize 2-layer bi-directional long short term memory (LSTM) with 128 neurons followed by a dropout layer and a fully-connected layer of 64 neurons. Again, to do classification, a softmax layer is used on top of the last dense layer (see Figure \ref{fig:rnn}).
\item \textbf{RNN-Attention} we added an attention layer on top an RNN model analogous to the one mentioned above to put more emphasis on the important waves of the input signal(s) that have most contribution in detecting the arrhythmia (see Figure \ref{fig:rnn-att}). The attention layer assigns a probability value to each feature vector extracted from the input by the RNN. In the probability vector, each value is the importance of the corresponding feature vector. Then, an expected value (i.e., it is a linear weighted vector) of the input feature vectors is computed according to  the weights provided by the attention layer. Finally, the weighted vector is fed into a softmax layer to perform the classification task.
\end{itemize}

We followed the aforementioned steps of the ELP pipeline in Section \ref{sec:methodology_1} in which we converted each input signal to an integer encoded vector and computed its corresponding embedding vector using a shallow neural network. Then, we used the embedding vectors as input for the mentioned models (i.e., the CNN, RNN and RNN-Attention) for building the detective models. 

\textbf{Implementation details} 
We trained all models with a maximum number of 25 epochs and batch size of 64 samples. The Adam optimizer was applied to minimize the loss with a learning rate $\alpha = 0.001$. To mitigate the effect of the overfitting problem, an $L_2$ regularization with a coefficient $\beta= 1e-5$  and a dropout approach with a probability of keeping input units of $0.8$ were used. We implemented the models using Python programming language and Google Tensorflow deep learning library on a machine equipped with 8 CPUs (Intel(R) Xeon(R) CPU $@$ 3.60 GHz), 32 GB memory
and Ubuntu 18.04. Also, in all experiments, we reported the best performance.
\subsection{Results and discussion}
We report the performance of all built models using all databases and show that following the ELP steps lead to better performance compared to the existing algorithms. We evaluate the models in terms of the overall accuracy, precision (positive predictive value (PPV)), recall (sensitivity), specificity, and F1-score. We also computed macro-averaging of F1-score (MF1) which is the sum of per-class F1-score over the number of classes.

We used a ten-fold cross-validation to assess the performance of the proposed method for the heartbeat classification task using the MIT-BIH arrhythmia database. Table \ref{tab:comparemitbih} presents the detection scores on the MIT-BIH arrhythmia database. We see that ELP work with the CNN, RNN and RNN-Attention approaches outperforms all other methods listed in the table. The RNN-Attention model performs as good as the CNN model indicating the attention mechanism helps in getting better performance. Furthermore, Table \ref{tab:cmmitbiharr} reports a confusion matrix of classified heartbeats and performance of each class achieved by the ELP while we use the CNN approach to build the classifier. According to Table \ref{tab:cmmitbiharr}, the smallest sensitivity values are obtained for the categories F and S. The reason is the class imbalance problem existed in the database where the group F has only 802 heartbeats and the group S has 2,777 heartbeats. An imbalanced dataset can negatively affect the performance of an machine learning algorithm. Typically, generating synthetic data or tweaking loss functions are used to mitigate this problem \cite{mousavi2019sleepeegnet}.

\begin{table} [htb] 
\caption{Comparison of performance of the proposed approach against other existing algorithms on the MIT-BIH arrhythmia database.}
 \centering{
\label{tab:comparemitbih}
\resizebox{0.8\linewidth}{!}{  
\begin{tabular}{lccc}
\toprule[\heavyrulewidth]\toprule[\heavyrulewidth]
\textbf{Work} &   \textbf{Approach}& \textbf{Accuracy (\%)} \\
\midrule
\textbf{ELP}& \textbf{CNN}& \textbf{97.00}   \\
\textbf{ELP}& \textbf{RNN}& \underline{96.96}   \\
\textbf{ELP}& \textbf{RNN-Attention}& \textbf{97.00}   \\
Kachuee et al. \cite{kachuee2018ecg}& Deep residual CNN & 93.4   \\
Acharya et al. \cite{acharya2017deep}& Augmentation + CNN & 93.47   \\
Li et al. \cite{li2016ecg}& DWT + random forest & 94.61   \\
Martis et al. \cite{martis2013application}& DWT + SVM & 93.8   \\

 \bottomrule  
 \multicolumn{3}{l}{DWT: Discrete wavelet transform; SVM: Support vector machine}
\end{tabular} }
}
\end{table}

\begin{table} [ht]  
\caption{ Confusion matrix and per-class performance achieved by the proposed method across all ten-folds using the CNN model and based on the MIT-BIH arrhythmia database.}
\renewcommand{\arraystretch}{1.4}
 \centering{
\label{tab:cmmitbiharr}
	\resizebox{1.0\linewidth}{!}{  
\begin{tabular}{ccccccccccc}
 \toprule
\textbf{} &  \textbf{} &  \multicolumn{5}{c} {Predicted} & \multicolumn{4}{c} {Per-class Performance (\%)} \\
 \cmidrule(lr){3-7} 
 \cmidrule(lr){8-11}
\textbf{} &  \textbf{}& N & S& V&F&Q & \textit{acc} & \textit{ppv}&\textit{sen}&\textit{spec} \\
\cline{1-11}
\multirow{5}{*}{Actual}
\textbf{} & \multicolumn{1}{l|}{N}
& 89774  & 203  &  357  &  37 &  91 & 97.35& 97.60 &99.24 &88.30 
 \\ 
\textbf{} & \multicolumn{1}{l|}{S}
& 757 & 1945  & 56  &  1  & 18 & 98.99 & 87.89 &70.04 &99.75 
 \\ 
\textbf{} &  \multicolumn{1}{l|}{V}
& 632  & 51 & 6449 & 44 &  47 & 98.77 & 91.88 &89.28 &99.94 
 \\ 
\textbf{} & \multicolumn{1}{l|}{F}
&  175    & 3  & 95 & 527  &   2 & 99.67 & 86.39 &65.71 &99.92 
 \\ 
\textbf{} & \multicolumn{1}{l|}{Q}
& 639   & 11   & 62   &   1  & 7314 & 99.20 & 97.89 &91.12 &99.84 
 \\ 
  \bottomrule 
\multicolumn{11}{l}{acc: accuracy; ppv: positive predictive value; sen: sensitivity; spec: specificity}
\end{tabular} }
}
\end{table}

We  employed  a  five-fold  cross-validation  to  evaluate  the  performance  of  the  proposed  method  for the atrial fibrillation classification task using the PhysioNet Computing in Cardiology
Challenge 2017 dataset. Table \ref{tab:compareafib17} shows a performance comparison of the 3 models (i.e., the CNN, RNN and RNN-Attention) following our proposed method on detecting atrial fibrillation against the state-of-the-art algorithms. From the table, we can observe that the ELP work with the CNN approach outperforms other methods listed in Table \ref{tab:compareafib17}, obtaining an MF1 score of 64.40\%. As it is shown in the table, the RNN-attention achieves better performance compared to the RNN, showing the attention mechanism leads to performance improvement. Applying the attention approach to the RNN (or other deep learning models) not only improves the model performance but also it provides interpretability into the model \cite{mousavi2020han,choi2016retain,mousavi2019ecgnet,mousavi2016learning}. Table \ref{tab:cmafib17} presents a confusion matrix and per-class performance of the atrial fibrillation classification task on the PhysioNet challenge AFIB dataset. Herein, we reported the model's results with the best performance (i.e., the CNN model). Even though the number of samples for class O (2,557) is larger than the number samples for class A (771), the model performs better for class A. This may be because the class \textit{Other rhythm} (O) contains a variety of rhythms with different morphologies that make it hard for the network to learn the associated patterns to the class O.  

\begin{table} [htb] 
\caption{Comparison of performance of the proposed approach against other algorithms for the atrial fibrillation (AFIB) classification task on the PhysioNet Computing in Cardiology
Challenge 2017 dataset.}
\renewcommand{\arraystretch}{1.4}
 \centering{
\label{tab:compareafib17}
	\resizebox{1.0\linewidth}{!}{  
\begin{tabular}{*{8}{c} }
\toprule[\heavyrulewidth]\toprule[\heavyrulewidth]
\textbf{Work}    & \textbf{Approach}  & \multicolumn{4}{c}{\textbf{Per-class Performance (F1\%)}}  & \multicolumn{2}{c}{\textbf{Overall Performance}}               \\    
\cmidrule(lr){3-6}
\cmidrule(lr){7-8}
\textbf{} &\textbf{}&  N&A&O&$\sim$&MF1 & Accuracy\\
\midrule
\multicolumn{1}{c}{\textbf{ELP}}   &   \textbf{CNN}   &   82.26  &   63.47  &   56.69  &  55.18  &  \textbf{ 64.40} & \underline{72.62} \\
\multicolumn{1}{c}{\textbf{ELP}}   &   \textbf{RNN}   &   79.88  &   56.06  &   44.32  &   43.31  &   55.89  &   67.66 \\
\multicolumn{1}{c}{\textbf{ELP}}   &   \textbf{RNN-Attention}   &   83.98  &  64.57  &   55.84  &  52.58  &  \underline{64.24}& \textbf{74.22}  \\
\multicolumn{1}{c}{Andreotti et al. \cite{andreotti2017comparing}}   &  Deep residual CNN   &  82.6  &   46.6  &  60.0  &  60.2  &   62.4 & - \\
 \bottomrule 
 \multicolumn{2}{l}{MF1: Macro-averaging of F1-score}
\end{tabular}}}
\end{table}

\begin{table} [ht]  
\caption{Confusion matrix and per-class performance achieved by the proposed method across all five-folds for the atrial fibrillation (AFIB) classification task on the PhysioNet Computing in Cardiology Challenge 2017 database.}
\renewcommand{\arraystretch}{1.4}
 \centering{
\label{tab:cmafib17}
	\resizebox{1.0\linewidth}{!}{  
\begin{tabular}{cccccccccc}
 \toprule
\textbf{} &  \textbf{} &  \multicolumn{4}{c} {Predicted} & \multicolumn{4}{c} {Per-class Performance (\%)} \\
 \cmidrule(lr){3-6} 
 \cmidrule(lr){7-10}
\textbf{} &  \textbf{}& N & A& O&$\sim$ & \textit{acc} & \textit{ppv}&\textit{sen}&\textit{spec} \\
\cline{1-10}
\multirow{4}{*}{Actual}
\textbf{} & \multicolumn{1}{l|}{N}
& 4221  & 53  &  738  &  63 & 78.65& 81.83 &83.17 &71.98 
 \\ 
\textbf{} & \multicolumn{1}{l|}{A}
& 70 & 463  & 207  &  18  & 93.75 & 66.05 &61.08 &96.93 
 \\ 
\textbf{} &  \multicolumn{1}{l|}{O}
& 839  & 172 & 1348 & 53 & 75.83 & 57.51 &55.89 &83.70 
 \\ 
\textbf{} & \multicolumn{1}{l|}{$\sim$}
&  57    & 13  & 51 & 157  & 97.01 & 53.95 &56.47 &98.37
 \\ 

  \bottomrule 
\multicolumn{10}{l}{\small acc: accuracy; ppv: positive predictive value; sen: sensitivity; spec: specificity}
\end{tabular} }
}
\end{table}

To evaluate the performance of our method for another AFIB classification task, we utilized a ten-fold cross-validation procedure on the MIT-BIH AFIB database with the ECG segment of size 5-s. Table \ref{tab:comparemitbihafib} reports the detection scores on the AFIB detection task. We see that the proposed work with the CNN model achieves a good performance but slightly low performance compared the Xia et al. \cite{xia2018detecting} work. 
\begin{table} [htb] 
\caption{Comparison of performance of the proposed approach against other state-of-the-art algorithms for the AFIB detection task on the MIT-BIH AFIB database with the ECG segment of size 5-s.}
\renewcommand{\arraystretch}{1.4}
 \centering{
\label{tab:comparemitbihafib}
	\resizebox{1\linewidth}{!}{  
\begin{tabular}{*{6}{c} }
\toprule[\heavyrulewidth]\toprule[\heavyrulewidth]
\textbf{Work}    & \textbf{Approach}  & \multicolumn{4}{c}{\textbf{Best Performance (\%)}}  \\    
\cmidrule(lr){3-6}
\textbf{} &\textbf{}&  \textit{accuracy}&\textit{ppv}&\textit{sensitivity}&\textit{specificity}\\
\midrule
\multicolumn{1}{c}{\textbf{ELP}}   &   \textbf{CNN}   &   \underline{98.17}  &  \textbf{97.78}  &   \underline{98.57}  &  {97.76}   \\
\multicolumn{1}{c}{\textbf{ELP}}   &   \textbf{RNN}   &   97.93  &   {97.63}  &  98.24  &  97.61  \\
\multicolumn{1}{c}{\textbf{ELP}}   &   \textbf{RNN-Attention}   &   97.96 &  \underline{97.87}  &   98.08  & \underline{97.84}  \\
\multicolumn{1}{c}{Xia et al. \cite{xia2018detecting}}   &  SWT + CNN   &  \textbf{98.63}  &   -  & \textbf{98.79}  &  \textbf{97.87}  \\
\multicolumn{1}{c}{Asgari et al. \cite{asgari2015automatic}}   &  SWT + SVM   &  -  &  - &  97.00  &  97.10  \\
\multicolumn{1}{c}{Jiang et al. \cite{jiang2012high}}   & \makecell{RR interval
irregularity + \\ P-wave absence }  &  -  &  - &  98.20  &  97.50  \\

 \bottomrule 
\multicolumn{6}{l}{ppv: positive predictive value; SWT: stationary wavelet transform}
\end{tabular}}}
\end{table}

From all experiments for three database, we can see that our method can result in better performance or comparable performance with smaller neural networks compared to other deep neural networks and existing algorithms. Therefore, this make the proposed method implementable on the devices with a limited hardware such as wearable devices.  It is worth mentioning that in the first step of the ELP pipeline (\nameref{step_1}), we extracted the waves in the ECG signals based on the extracted R-peaks and employing adaptive searching windows, and used a K-means clustering algorithm to cluster waves to build the vocabulary. We believe, applying better segmentation algorithms or more sophisticated clustering methods can yield to higher detection scores.

\section{Conclusion}
\label{sec:conclu}
In this study, we proposed a new technique to analyze ECG signals named ECG language processing (ELP). The proposed approach is composed of two main steps: 1) Creating a Wave Vocabulary, building a vocabulary of waves based on the extracted waves from the ECG signals, and 2) Training and Deploying, developing predictive and detective models using the extracted vocabulary and machine learning algorithms for different clinical tasks. The experiment results on two different tasks, including the heartbeat classification and atrial fibrillation tasks with three databases show that our method results in the state-of-the-art performance. Future work includes, but not limited to, improving the segmentation and creating the vocabulary steps to improve the  performance of the detection process and applying the ELP method for other biomedical applications such as the prediction of arrhythmia (see Section \ref{sec:elpexamples} for more examples).  



\section*{Acknowledgment}
This study is based upon work supported by the National Science Foundation under Grant Number 1657260. Research reported in this publication was supported by the National Institute On Minority Health And Health Disparities of the National Institutes of Health under Award Number U54MD012388.


\ifCLASSOPTIONcaptionsoff
  \newpage
\fi



%
\bibliographystyle{IEEEtran}
\bibliography{IEEEabrv,bare_jrnl}




%








\end{document}